\documentclass[twocolumn,showpacs,preprintnumbers,prb]{revtex4}
\usepackage{amsmath}
\usepackage{graphicx}
\usepackage{dcolumn}
\usepackage{bm}

\begin{document}
\title{Symmetry-adapted Wannier functions
in the maximal localization procedure}
\author{R. Sakuma}
\affiliation{Division of Mathematical Physics,
Lund University, S\"{o}lvegatan 14A 223 62
Lund, Sweden}
\date{\today }

\begin{abstract}
A procedure
to construct symmetry-adapted Wannier functions
in the framework of the maximally-localized
Wannier function approach[Marzari and Vanderbilt,
Phys. Rev. B \textbf{56}, 12847 (1997); Souza, Marzari,
and Vanderbilt, \textit{ibid.} \textbf{65}, 035109 (2001)]
is presented.
In this scheme the minimization of
the spread functional of the Wannier functions is performed
with constraints that are derived from
symmetry properties of
the specified
set of the Wannier functions and
the Bloch functions used to construct them, therefore
one can obtain a solution that does not necessarily
yield the global minimum of the spread functional.
As a test of this approach,
results of atom-centered Wannier functions for
GaAs and Cu are presented.
\end{abstract}

\pacs{71.15.Ap}
\maketitle
% 71.15.Ap Basis sets (LCAO, plane-wave, APW, etc.)
% and related methodology
% (scattering methods, ASA, linearized methods, etc.)

\section{introduction}
After its proposal by Marzari and Vanderbilt\cite{Marzari97},
the maximally-localized Wannier function
approach\cite{Marzari97,Souza01,Marzari12}
has been widely used
as a convenient tool to construct
localized orthonormal functions in crystals.
These Wannier functions are
obtained from unitary transformation of the Bloch functions,
whose phase factors are chosen so that
the spatial spreads of
the Wannier functions\cite{Blount} are minimized.
The maximally-localized Wannier functions have been employed
for a number of applications, such as analysis of 
chemical bonding and basis functions
for linear-scaling calculations or 
model Hamiltonians in strongly-correlated
systems\cite{Marzari12}.

As found by Souza \textit{et al.}\cite{Souza01}
in the case of the $s$-like Wannier function of Cu,
the Wannier functions
obtained in the maximal-localization procedure
do not necessarily
reflect the spatial symmetry of the system;
the center of the 4$s$-like Wannier function of Cu is not
at the Cu atom but at the tetrahedral interstitial site.
A similar result was reported
by Thygesen \textit{et al.}\cite{Thygesen05}.
This feature of the maximal-localization approach
has some drawbacks;
the center of the maximally-localized
Wannier functions is not necessarily
on atom positions
or other high-symmetry points, which sometimes
makes the interpretation
of the obtained Wannier functions difficult.
Furthermore, due to
the lack of definite symmetry in the Wannier functions,
one has to calculate the transformation matrix from the Bloch
functions to the Wannier functions for all $\mathbf{k}$-points
in the first Brillouin zone,
not only inside the irreducible part of it.

The connection between the symmetry of the crystal
and the properties of Wannier functions was first discussed by
des Cloizeaux\cite{Cloizeaux63} %,Cloizeaux64a,Cloizeaux64b}
from the view point of group theory.
His basic idea is
that the Wannier functions can be chosen to be
the basis of the irreducible representations of a subgroup of
the full symmetry group of the system whose elements 
do not
change the given point in the unit cell, and
he derived the relation between
these Wannier functions and
the eigenfunctions of the one-particle
Hamiltonian of the crystal(i.e. Bloch states).
This idea is based on the
site-symmetry group and the theory of
the induced representations\cite{Evarestov97},
and there have been a number of works
considering these symmetry-adapted Wannier
functions\cite{Kohn73,Boehm79,Kruger87, Sporkmann94,
Evarestov97,Smirnov01,Smirnov05,Posternak02,Casassa06}.
Based on this idea, in this work we propose
a procedure to construct symmetry-adapted Wannier functions
in the framework of the maximally-localized Wannier
function approach.
Considering the symmetry properties of the specified
set of the Wannier functions
and the Bloch functions used,
we derive a formula that the transformation matrix
follows, and perform
the minimization of the spread functional
with this symmetry constraint.
Our procedure
enables one to control the symmetry and center of
the Wannier functions,
and 
it also enables one to generate the transformation matrix
for a general $\mathbf{k}$-point from
its symmetry-equivalent point inside the irreducible
Brillouin zone(IBZ),
which simplifies
the minimization of the spread functional.
As a test of our approach, we consider
Wannier functions of GaAs and Cu. %, the materials considered in
%the original papers\cite{Marzari97,Souza01}.
\section{method}
\subsection{Symmetry-adapted Wannier functions}
\subsubsection{Site symmetry group and symmetry-adapted
Wannier functions}
In this section
we summarize the main points of site symmetry group
and symmetry-adapted Wannier functions
following Ref.~\onlinecite{Evarestov97}.
We refer to Refs.~\onlinecite{Cloizeaux63,Evarestov97}
for details of the theory.

The starting point of this idea is to specify a set of
positions in real space(``sites'') in which one or more
Wannier functions will
be centered.
These sites can be either atomic positions, at chemical
bonds,
or interstitial sites, depending on the case of interest.
The site symmetry group of a given point
$\mathbf{q}$, denoted by $G_{q}$,
is a subgroup of 
%are the elements of the full symmetry group of the crystal $G$
the full symmetry group of the crystal $G$
whose elements leave $\mathbf{q}$ unchanged:
namely, $g_{q}=(R_{q}|\mathbf{v}_{q}+\mathbf{T}_{q}) \in
G_{q}$ satisfies
\begin{align}
g_{q}\mathbf{q}&=
(R_{q}|\mathbf{v}_{q}+\mathbf{T}_{q})
\mathbf{q} \nonumber\\
&= 
R_{q}\mathbf{q}+\mathbf{v}_{q}+
\mathbf{T}_{q} =
\mathbf{q},
\end{align}
where $R_{q}, \mathbf{v}_{q}+\mathbf{T}_{q}$ are
the rotation and
the translation part of the symmetry operation with
$\mathbf{T}_{q}$ a lattice translation vector.
The full symmetry group $G$ can be decomposed into
left cosets of the subgroup $G_{q}$ as
\begin{equation}
G=\sum_{j,n}g_{jn}G_{q},
\end{equation}
where
\begin{equation}
g_{jn}=(R_{j}|\mathbf{v}_{j}+\mathbf{T}_{j}+
\mathbf{T}_{n}).
\end{equation}
In the above equation, 
$g_{j0}$ is one of the symmetry operations that maps
$\mathbf{q}$ to
its symmetry-equivalent point $\mathbf{q}_{j}$ as
\begin{align}
\mathbf{q}_{j}& \equiv
g_{j 0} \mathbf{q} \nonumber\\
&=(R_{j}|\mathbf{v}_{j}+\mathbf{T}_{j})\mathbf{q}
\nonumber \\
&=
R_{j}\mathbf{q}+\mathbf{v}_{j}+\mathbf{T}_{j}.
\end{align}
Here $j=1$ corresponds to the original point $\mathbf{q}$
(i.e. $\mathbf{q}_{1}=\mathbf{q}$ and $g_{1 0}=(E|\mathbf{0})$,
where $E$ denotes the identity operation).
These points $\{ \mathbf{q}_{j} \}$ constitute
a crystallographic orbit whose multiplicity is
% the number of $\mathbf{q}_{j}$ (including
%$\mathbf{q}$) is
given by $n_{G}/n_{G_{q}}$,
where $n_{G}$ is the number of symmetry
operations in the full crystal group
without pure translations,
and $n_{G_{q}}$ is the number of elements in $G_{q}$.
The vector $\mathbf{T}_{j}$ is chosen so that
$g_{j0}$ transforms $\mathbf{q}$
to the point $\mathbf{q}_{j}$
which lies in the unit cell.
%(primitive cell or Wigner-Seitz cell).

From the site symmetry group for a given point $\mathbf{q}$,
the symmetry-adapted Wannier functions centered at
$\mathbf{q}$  are defined
as the basis functions of the
irreducible representations of the site symmetry
group $G_{q}$; these Wannier functions are represented as
$W_{i1}^{(\beta)}(\mathbf{r})
\equiv
W^{(\beta)}_{i}(\mathbf{r}-\mathbf{q}_{1})$, where
$\beta$ labels the irreducible representations and
$i=1,2,...,n_{\beta}$ runs over the basis functions of
the irreducible representation $\beta$, and
$n_{\beta}$ is the dimension of the
irreducible representation $\beta$.
For $g_{q} \in G_{q}$
these Wannier functions transform as
% $W^{(\beta)}_{i1}(\mathbf{r})$
%transforms as
\begin{align}
\hat{g}_{q}W^{(\beta)}_{i1}(\mathbf{r})&=
W^{(\beta)}_{i}(g^{-1}_{q}\mathbf{r} - \mathbf{q}) \nonumber\\
&=W^{(\beta)}_{i}(R^{-1}_{q}(\mathbf{r}-\mathbf{v}_{q}
-\mathbf{T}_{q}-R_{q}\mathbf{q})) \nonumber\\
&=
W^{(\beta)}_{i}(R_{q}^{-1}(\mathbf{r}-\mathbf{q})) \nonumber\\
&=
\sum_{i' = 1}^{n_{\beta}}
d^{(\beta)}_{i' i}(R_{q})
W^{(\beta)}_{i' 1}(\mathbf{r}),
\label{eq:gqW}
\end{align}
%where $n_{\beta}$ is the dimension of the irreducible representation $\beta$.
From $W_{i1}^{(\beta)}(\mathbf{r})$
we can generate Wannier functions
centered at $\mathbf{q}_{j}$ as
\begin{align}
W_{ij}^{(\beta)}(\mathbf{r})
& \equiv
\hat{g}_{j0}W_{i1}^{(\beta)}(\mathbf{r})\nonumber\\
&=
W_{i}^{(\beta)}(R_{j}^{-1}(\mathbf{r}-\mathbf{q}_{j}))
\end{align}
%% The functions centered in other unit cells are generated through
%% \begin{align}
%% W_{ij}^{(\beta)}(\mathbf{r}-\mathbf{T}_{n})&=
%% ( E|\mathbf{T}_{n} )
%% W_{ij}^{(\beta)}(\mathbf{r}) \nonumber\\
%% &=
%% W_{i}^{(\beta)}(R_{j}^{-1}(\mathbf{r}-\mathbf{q}_{j}-\mathbf{T}_{n}))
%% \nonumber\\
%% &=
%% ( R_{j}|\mathbf{v}_{j}+\mathbf{T}_{j}+\mathbf{T}_{n} )
%% W_{i1}^{(\beta)}(\mathbf{r}) \nonumber\\
%% &=
%% \hat{g}_{jn}W_{i1}^{(\beta)}(\mathbf{r})
%% \end{align}
%% It can be shown that
%% for a given
%% $g_{jn}$,
%% there exist for all elements
%% of the full symmetry group, $g=(R|\mathbf{v})\in G$,
%% one pair of $g_{q}\in G_{q}$ and $g_{j' 0}$
%% that satisfies the following equation:
%% \begin{align}
%% g &=
%% (E|\mathbf{T}_{jj'}+R\mathbf{T}_{n})
%% g_{j'\mathbf{0}}
%% \,
%% g_{q} \,
%% g_{j n}^{-1},
%% \end{align}
%% where
%% \begin{align}
%% R&=R_{j'}R_{q}R_{j}^{-1} \\
%% \mathbf{T}_{jj'}&=g\mathbf{q}_{j}
%% -\mathbf{q}_{j'}.
%% \end{align}
%% The functions $W_{ij}^{(\beta)}(\mathbf{r})$ form
%% a (reducible) basis of
%% $G$:
%% \begin{align}
%% \hat{g} W_{ij}^{(\beta)}(\mathbf{r}-\mathbf{T}_{n})
%% &=
%% \sum_{i'}
%% d_{i' i}^{(\beta)}(R_{j'}^{-1}R R_{j}) \nonumber\\
%% &W_{i' j'}^{(\beta)}
%% (\mathbf{r}-\mathbf{T}_{jj'}-R\mathbf{T}_{n}).
%% \label{eq:gW}
%% \end{align}
Therefore the symmetry-adapted
Wannier functions can be specified
by one representative point of their centers
(i.e. Wyckoff position)
and the irreducible representations of the corresponding
site-symmetry group.
If the irreducible representation
$d^{(\beta)}_{i'i}$ is real, the corresponding
Wannier functions can be
chosen to be real\cite{Cloizeaux63}.
From these symmetry-adapted Wannier functions
$W_{ij}^{(\beta)}(\mathbf{r})$, one can
construct the Bloch functions
$\psi_{\mathbf{k} ij}^{(\beta)}(\mathbf{r})$ as
\begin{equation}
\psi^{(\beta)}_{\mathbf{k} ij}(\mathbf{r})=
\sum_{n}
e^{i\mathbf{k}\cdot\mathbf{T}_{n}}
W_{ij}^{(\beta)}(\mathbf{r}-\mathbf{T}_{n}),
\label{eq:qBloch}
\end{equation}
\begin{equation}
W_{ij}^{(\beta)}(\mathbf{r}-\mathbf{T}_{n}) =
\frac{1}{N_{\mathbf{k}}}
\sum_{\mathbf{k}}
e^{-i\mathbf{k}\cdot\mathbf{T}_{n}}
\psi^{(\beta)}_{\mathbf{k}ij}(\mathbf{r}).
\end{equation}
Here $N_{\mathbf{k}}$ is the number of
$\mathbf{k}$-points.

To understand how these Wannier functions transform
with respect to the operations in the full symmetry group
of the system,
the theory of induced representations is used;
it can be shown that, for a given $g_{jn}$
and any element of the full symmetry group,
$g=(R|\mathbf{v})\in G$,
there exist
one pair of $g_{q}\in G_{q}$ and $g_{j' 0}$
that satisfy the following equation\cite{Evarestov97}:
\begin{align}
 g &=
 (E|\mathbf{T}_{j'j}+R\mathbf{T}_{n})
 g_{j'0}
 \,
 g_{q} \,
 g_{j n}^{-1},
\label{eq:gdecomp}
 \end{align}
% where $E$ denotes the identity operation and
where
 \begin{align}
 R_{j'}R_{q}R_{j}^{-1}&= R\\
 \mathbf{T}_{j'j}&=g\mathbf{q}_{j}
 -\mathbf{q}_{j'}.
\end{align}
Using Eq.~(\ref{eq:gdecomp}),
it can be shown that
these symmetry-adapted Wannier functions and
the Bloch functions transform as\cite{Evarestov97}
\begin{align}
\hat{g}
W_{ij}^{(\beta)}(\mathbf{r}-\mathbf{T}_{n})
&=
\sum_{i'=1}^{n_{\beta}}
d_{i' i}^{(\beta)}(R_{j'}^{-1}R R_{j}) \nonumber\\
&\quad
\times W_{i' j'}^{(\beta)}
(\mathbf{r}-\mathbf{T}_{j'j}-R\mathbf{T}_{n}),
\label{eq:Wannier}
\end{align}
\begin{align}
\hat{g}
\psi^{(\beta)}_{\mathbf{k}ij}(\mathbf{r})&=
e^{-i R\mathbf{k}\cdot \mathbf{T}_{j'j}}
\sum_{i'=1}^{n_{\beta}}
d_{i' i}^{(\beta)}(R_{j'}^{-1}R R_{j})
\psi_{R\mathbf{k} i'j'}^{(\beta)}(\mathbf{r}).
\label{eq:Bloch}
\end{align}
In the above equations the index $j'$ or the symmetry operation
$g_{j'}$ is determined by $g$ and $g_{j}$ according to
Eq.~(\ref{eq:gdecomp}).
By writing $I=\{ij(\beta)\}$, Eq.~(\ref{eq:Bloch}) can be rewritten as
\begin{align}
\hat{g}
\psi_{\mathbf{k}I}(\mathbf{r})&=
\sum_{I'} D_{I' I}(g,\mathbf{k})
\psi_{R\mathbf{k} I'}(\mathbf{r}),
\label{eq:Bloch2}
\end{align}
where $D(g,\mathbf{k})$ is a block diagonal matrix
\begin{align}
D_{i'j'\beta',ij\beta}(g,\mathbf{k})&=
\delta_{\beta' \beta}
e^{-iR\mathbf{k}\cdot\mathbf{T}_{j' j}}
d^{(\beta)}_{i' i}(R^{-1}_{j'} R R_{j}).
\label{eq:DRk}
\end{align}
Due to its block-diagonal form,
the matrix $D(g,\mathbf{k})$ can contain blocks corresponding
to nonequivalent Wannier centers(different Wyckoff positions), and
the number of the blocks in $D(g,\mathbf{k})$ is given as
the sum of the number of
the irreducible representations considered
for a given set of Wannier centers;
when there are more than one set of Wannier functions
belonging to the same irreducible representation,
$D(g,\mathbf{k})$ contains the same number of
identical blocks as the number of these multiple sets.

\subsubsection{Construction of the Bloch functions from the
eigenstates of the Hamiltonian}
Our objective is to construct the Bloch wavefunctions
defined in Eq.~(\ref{eq:qBloch}), which are related to
the symmetry-adapted
Wannier functions through unitary transformation and
transform according to Eq.~(\ref{eq:Bloch2}),
from the linear combination of
the eigenfunctions of some
one-particle Hamiltonian that is invariant under
the full symmetry operations of the system. In
this work we use
the Kohn-Sham Hamiltonian of density functional
theory\cite{DFT}. By using the Kohn-Sham wavefunctions
$\psi^{\textrm{KS}}_{\mathbf{k}\mu}(\mathbf{r})$, we construct
the orthonormal Bloch functions as
\begin{equation}
\psi_{\mathbf{k}I}(\mathbf{r})=
\sum_{\mu} U_{\mu I}(\mathbf{k})
\psi^{\textrm{KS}}_{\mathbf{k} \mu}(\mathbf{r}).
\label{eq:IKS}
\end{equation}
Since the Kohn-Sham wavefunctions $\psi^{\textrm{KS}}_{\mathbf{k}\mu}(\mathbf{r})$
form the basis of the irreducible representations
of the full symmetry group of the system $G$, they transform as
\begin{align}
\hat{g} \psi^{\textrm{KS}}_{\mathbf{k}\mu}(\mathbf{r}) &=
\sum_{\mu'}\tilde{d}_{\mu' \mu}(g,\mathbf{k})
\psi^{\textrm{KS}}_{R\mathbf{k}\mu'}(\mathbf{r})
\label{eq:KStrafo}
\end{align}
for $g=(R|\mathbf{v}) \in G$.
From Eqs. (\ref{eq:Bloch2}), (\ref{eq:IKS}), (\ref{eq:KStrafo}),
one can obtain the following relation
between the transformation matrices $U(\mathbf{k})$ and
$U(R\mathbf{k})$:
\begin{equation}
U(R\mathbf{k}) D(g,\mathbf{k})
=
\tilde{d}(g,\mathbf{k}) U(\mathbf{k})
\label{eq:URk}
\end{equation}
Therefore $U(R\mathbf{k})$ can be calculated
 from $U(\mathbf{k})$
by providing $\tilde{d}(g,\mathbf{k})$ and $D(g,\mathbf{k})$.
In this work
we do not consider
time-reversal symmetry, but Eq.~(\ref{eq:URk})
can be generalized to include it.
For the symmetry
operations that transform $\mathbf{k}$ to itself,
namely, for the operations in the little group of $\mathbf{k}$
denoted by $G_{\mathbf{k}}$,
Eq.~(\ref{eq:URk})
yields the condition that $U(\mathbf{k})$
has to fulfill:
\begin{equation}
U(\mathbf{k})=
\tilde{d}(g_{\mathbf{k}},\mathbf{k}) U(\mathbf{k})
D^{\dagger}(g_{\mathbf{k}},\mathbf{k})
\quad g_{\mathbf{k}} \in G_{\mathbf{k}},
\label{eq:Uk}
\end{equation}
where unitarity of $D(g,\mathbf{k})$ is used.

Equations (\ref{eq:Wannier}), (\ref{eq:Bloch}),
(\ref{eq:URk}), and (\ref{eq:Uk}) are
the central equations
of this work. We force the transformation matrix
$U(\mathbf{k})$ to follow Eq.~(\ref{eq:URk})
for all $g =(R|\mathbf{v}) \in G$, which guarantees
that the resulting Wannier functions transform according
to Eq.~(\ref{eq:Wannier}).
Since $U(\mathbf{k})$ and $U(R\mathbf{k})$
are related by Eq.~(\ref{eq:URk}), we need
to calculate the
transformation matrix only for $\mathbf{k}$-points inside the IBZ,
which reduces the computational cost. Note that
since in practice the Wannier functions are calculated
using
a limited subspace spanned by a finite number of the Kohn-Sham
states inside a chosen ``energy window'',
it is not possible to construct $U(\mathbf{k})$ for
any desired irreducible representations.
If a given irreducible representation is not compatible
with the symmetry of the Kohn-Sham states
inside the energy window,
Eq.~(\ref{eq:Uk}) cannot be fulfilled.

\subsection{Maximally-localized Wannier functions}
In the maximally-localized Wannier function
approach\cite{Marzari97,Souza01},
the Wannier functions are obtained
by minimizing the spread
functional
\begin{equation}
\Omega = \sum_{I}
\Bigl [
\langle
\mathbf{0}I|
\mathbf{r}^{2}
| \mathbf{0}I
\rangle
-
\langle
\mathbf{0}I|
\mathbf{r}
|\mathbf{0}I
\rangle^{2}
\Bigl ],
\label{eq:Omega}
\end{equation}
where $|\mathbf{0}I\rangle=|W_{\mathbf{0} I}\rangle$ is the
Wannier function $I$ whose center
is in the cell $\mathbf{T}_{n}=\mathbf{0}$.
The matrix elements
$\langle
\mathbf{0}I|
\mathbf{r}|
\mathbf{0}I
\rangle$ and
$\langle
\mathbf{0}I|
\mathbf{r}^{2}
|\mathbf{0}I
\rangle$ are calculated as\cite{Marzari97,Blount}
\begin{align}
\langle
\mathbf{0}I|
\mathbf{r}
| \mathbf{0}I
\rangle &= \frac{i}{N_{\mathbf{k}}}\sum_{\mathbf{k},\mathbf{b}}
\textrm{w}_{b}
\mathbf{b}
\Bigl [
\langle u_{\mathbf{k}I}|u_{\mathbf{k}+\mathbf{b}I} \rangle
-1
\Bigr ]
\label{eq:r}
 \\
\langle
\mathbf{0}I|
\mathbf{r}^{2}
| \mathbf{0}I
\rangle &= \frac{1}{N_{\mathbf{k}}}
\sum_{\mathbf{k},\mathbf{b}}
\textrm{w}_{b}
\Bigl [
2
-
2
\textrm{Re}
\langle u_{\mathbf{k}I}|u_{\mathbf{k}+\mathbf{b}I} \rangle
\Bigr ]
\label{eq:r2}
\end{align}
where $u_{\mathbf{k}I}(\mathbf{r})$ is the cell-periodic part
of the Bloch function (Eq.~(\ref{eq:IKS}))
and
$\mathbf{b}$ is the vector that connects a given $\mathbf{k}$-point with
its neighbors and $\textrm{w}_{b}$ is its weight.
The spread functional given by Eq.~(\ref{eq:Omega})
can conveniently be  decomposed as
$\Omega = \Omega_{\textrm{I}}+\Omega_{\textrm{OD}}+
\Omega_{\textrm{D}}$, where\cite{Marzari97}
\begin{align}
\Omega_{\textrm{I}} &= \sum_{I}
\Bigl [
\langle
\mathbf{0}I|
\mathbf{r}^{2}
|\mathbf{0}I
\rangle
-
\sum_{\mathbf{T}_{n} I'}
|\langle
\mathbf{T}_{n} I'|
\mathbf{r}
| \mathbf{0} I
\rangle|^{2}
\Bigr ],
\label{eq:Omega_I}
\\
\Omega_{\textrm{OD}} &= \sum_{I \ne I'}
\sum_{\mathbf{T}_{n}}
|
\langle
\mathbf{T}_{n}I'|
\mathbf{r}
|\mathbf{0}I
\rangle
|^{2},
\label{eq:Omega_OD}
\\
\Omega_{\textrm{D}} &= \sum_{I}
\sum_{\mathbf{T}_{n} \ne \mathbf{0}}
|
\langle
\mathbf{T}_{n}I|
\mathbf{r}
|\mathbf{0}I
\rangle
|^{2}.
\label{eq:Omega_D}
\end{align}
It can be shown that $\Omega_{\textrm{I}}$ is invariant under
the unitary transformation
of the Bloch functions\cite{Marzari97}.
The algorithm to minimize the spread functional is given
in Refs.~\onlinecite{Marzari97} and \onlinecite{Souza01}
for both the cases where the bands of interest
are isolated from other bands and 
are entangled with other bands.
The minimization can be done as a postprocess to
density-functional calculations, and the necessary input
data are
$\langle u_{\mathbf{k}\mu'}^{\textrm{KS}}|
u_{\mathbf{k}+\mathbf{b}\mu}^{\textrm{KS}}\rangle
$,
the overlap matrix elements between the states
at $\mathbf{k}$ and $\mathbf{k}+\mathbf{b}$ from which
the spread functional is calculated via Eqs.~(\ref{eq:r})
and (\ref{eq:r2}),
and
the initial guess of the transformation matrix, which is
obtained
by orthonormalizing the following matrix\cite{Marzari97}
\begin{equation}
A_{\mu I}(\mathbf{k})=
\langle \psi^{\textrm{KS}}_{\mathbf{k}\mu}
|
w_{I}\rangle,
\label{eq:amn}
\end{equation}
where $w_{I}$ is an initial guess of the Wannier function
$I$.

\subsection{Minimization of the spread functional
under symmetry constraint}
\subsubsection{Input data}
To perform minimization of the spread functional under
the symmetry constraint (Eqs.~(\ref{eq:URk}) and (\ref{eq:Uk})),
in addition to the overlap matrix elements
$\langle u^{\textrm{KS}}_{\mathbf{k} \mu'} | u^{\textrm{KS}}_{\mathbf{k}+\mathbf{b} \mu} \rangle$
and the initial guess of the
transformation matrices(Eq.~(\ref{eq:amn})),
one needs
the matrix representation of the symmetry operations
in the basis of the Bloch functions defined
in Eq.~(\ref{eq:qBloch}) and the Kohn-Sham states,
$D_{I'I}(g,\mathbf{k})$ (Eq.~(\ref{eq:DRk}))
 and
$\tilde{d}_{\mu' \mu}(g, \mathbf{k})$ (Eq.~(\ref{eq:KStrafo})).
%The former is obtained by specifying centers and
%irreducible representations of the Wannier functions.
The former is in many cases obtained
by specifying the center
and the character of the Wannier functions
(ex. $s$, $p$, $d$), and calculating the rotation matrix
for the elements of the corresponding site-symmetry group
expressed in the basis of these functions.
The latter is calculated from the Kohn-Sham wavefunctions as
\begin{equation}
\tilde{d}_{\mu' \mu}(g,\mathbf{k}) = 
\int \psi^{\textrm{KS}*}_{R\mathbf{k} \mu'}(\mathbf{r})
\psi^{\textrm{KS}}_{\mathbf{k} \mu}(g^{-1}\mathbf{r}) d^{3}r .
\label{eq:dks}
\end{equation}
Similar to the overlap matrices
$\langle u^{\textrm{KS}}_{\mathbf{k}\mu'}|
u^{\textrm{KS}}_{\mathbf{k}+\mathbf{b}\mu}\rangle$,
$\tilde{d}_{\mu' \mu}(g,\mathbf{k})$
in Eq.~(\ref{eq:dks}) can be calculated
with any basis set,
and after all data are calculated
the procedure is basis independent,
as in the original maximally-localized
Wannier function approach.
We also note that
this procedure does not require any specific phase factor
relation between $\psi_{\mathbf{k}\mu}^{\textrm{KS}}$ and
$\psi_{R\mathbf{k}\mu'}^{\textrm{KS}}$;
the Kohn-Sham wavefunctions at $R\mathbf{k}$ can
be calculated independently from those at $\mathbf{k}$,
or can be generated from the wavefunctions at $\mathbf{k}$
by performing symmetry operations.
It is also important to include all degenerate states
in the calculation of $\tilde{d}_{\mu'\mu}(g,\mathbf{k})$,
inside the specified energy window.

The initial transformation matrix
$U(\mathbf{k}) (\mathbf{k} \in \textrm{IBZ})$ has
to follow Eq.~(\ref{eq:Uk}).
We can construct $U(\mathbf{k})$ that fulfills this
requirement iteratively as follows; starting from 
$U(\mathbf{k})=U_{0}(\mathbf{k})$ that is calculated from
the initial guess of the Wannier
functions (Eq.~(\ref{eq:amn})),
we first calculate 
\begin{equation}
U'(\mathbf{k})=\frac{1}{N_{g_{\mathbf{k}}}}\sum_{g_{\mathbf{k}}}
\tilde{d}(g_{\mathbf{k}},\mathbf{k})
U_{0}(\mathbf{k})
D^{\dagger}(g_{\mathbf{k}},\mathbf{k}),
\end{equation}
and in the next step this $U'(\mathbf{k})$ is
orthonormalized by, e.g. using
singular value decomposition.
This cycle is repeated until we get converged $U(\mathbf{k})$.
For a limited energy window, it is not always possible to
construct $U(\mathbf{k})$ for a given set of the irreducible
representations.
A measure to check the convergence of $U(\mathbf{k})$ can be
\begin{equation}
\sum_{g_{\mathbf{k}}\in G_{\mathbf{k}}} 
||
\mathbf{1} - U^{\dagger}(\mathbf{k})
\tilde{d}(g_{\mathbf{k}},\mathbf{k})
U(\mathbf{k})
D^{\dagger}(g_{\mathbf{k}},\mathbf{k})
||,
\end{equation}
which is zero if $U(\mathbf{k})$ fulfills Eq.~(\ref{eq:Uk}).

\subsubsection{Isolated set of bands}
In the case where we construct
$N$ Wannier functions from $N$ bands that are separated
from all other bands, since any unitary transformation of
the Bloch states does not change $\Omega_{I}$,
we only have to
consider the variation of
$\tilde{\Omega}=\Omega_{D}+\Omega_{OD}$
with respect to the change
\begin{equation}
U_{\mu I}(\mathbf{k})
\to
\sum_{I'}
U_{\mu I'}(\mathbf{k})
\bigl [
\delta_{I' I}+dW_{I' I}(\mathbf{k})
\bigr ],
\end{equation}
where $dW(\mathbf{k})$
is an infinitesimal antiunitary matrix\cite{Marzari97}.
Using the relation Eq.~(\ref{eq:URk}),
for $\mathbf{k} \in \textrm{IBZ}$
the gradient of the spread functional is calculated as
\begin{align}
\frac{d \Omega}{d W(\mathbf{k})} &=
\frac{1}{n(\mathbf{k})}\sum_{g}
\frac{\partial \Omega}{\partial W(R\mathbf{k})}
\frac{\partial W(R\mathbf{k})}
{\partial W(\mathbf{k})} \nonumber\\
&=
\frac{1}{n(\mathbf{k})}\sum_{g}
D(g,\mathbf{k})
G^{(R\mathbf{k})}
D^{\dagger}(g,\mathbf{k}),
\end{align}
where $n(\mathbf{k})$
is the number of the
symmetry operations that leave $\mathbf{k}$
unchanged, and
$G^{(R\mathbf{k})}$ is the gradient of $\Omega$ with respect
to $W(R\mathbf{k})$,
whose explicit form is given in Ref.~\onlinecite{Marzari97}.
It can be shown the new set of $U(\mathbf{k})$ optimized
along
this direction also satisfy Eq.~(\ref{eq:Uk}).
The transformation matrices for $\mathbf{k}$-points 
not inside the IBZ are obtained via Eq.~(\ref{eq:URk}).

\subsubsection{Entangled bands}
When we construct the Wannier functions from the states
that are entangled with other bands,
generally
the number of the Bloch states inside
a given energy window is larger than
the number of the Wannier functions $N$.
Following Ref.~\onlinecite{Souza01}, in this case
we minimize the spread functional
using the two-step procedure;
first we determine the optimal subspace inside the specified
energy window spanned by $N$ orthonormal states
that minimizes $\Omega_{\textrm{I}}$,
and in the second
step the remaining part of the spread functional,
$\Omega_{\textrm{D}}+\Omega_{\textrm{OD}}$, is
minimized within the chosen subspace.
In the first step, we search for the optimal $N$
wavefunctions
\begin{equation}
\psi^{(opt)}_{\mathbf{k} I}(\mathbf{r})=
\sum_{\mu} U^{(opt)}_{\mu I}(\mathbf{k})
\psi^{\textrm{KS}}_{\mathbf{k} \mu}(\mathbf{r})
\label{eq:psiopt}
\end{equation}
which minimize $\Omega_{\textrm{I}}$ and also
transform according to Eq.~(\ref{eq:Bloch}).
This set of wavefunctions
are obtained by the variation of
$\Omega_{\textrm{I}}$
\begin{equation}
\frac{\delta}{\delta u^{(opt)*}_{\mathbf{k} I}}
\Bigl [
\Omega_{\textrm{I}}
-\sum_{\mathbf{k},I,I'}
\Lambda_{I'I}(\mathbf{k})
\langle
u^{(opt)}_{\mathbf{k} I}
|
u^{(opt)}_{\mathbf{k} I'}
\rangle
\Bigr ] = 0,
\end{equation}
where $\Lambda_{I'I}(\mathbf{k})$ is a Lagrange multiplier and
$u^{(opt)}_{\mathbf{k}I}$ is the periodic part of
$\psi^{(opt)}_{\mathbf{k}I}$.
In this work
we calculate $\psi^{(opt)}_{\mathbf{k} I}$
by using the steepest descend method; in each iteration
the wavefunctions are minimized along the direction
\begin{equation}
\delta \psi^{(opt)}_{\mathbf{k} I} =
\tilde{Z}(\mathbf{k})
\psi^{(opt)}_{\mathbf{k} I} - \sum_{I'} \lambda_{I'I}(\mathbf{k})
\psi^{(opt)}_{\mathbf{k} I'},
\label{eq:sdvec}
\end{equation}
where $\tilde{Z}(\mathbf{k})$ is
the Hermitian operator defined as
\begin{align}
\tilde{Z}_{\mu \mu'}(\mathbf{k})
&=
\langle u^{\textrm{KS}}_{\mathbf{k} \mu}|
\tilde{Z}(\mathbf{k})
| u^{\textrm{KS}}_{\mathbf{k} \mu'} \rangle \nonumber\\
&=
\Bigl [
\frac{1}{n(\mathbf{k})}
\sum_{g} \tilde{d}^{\dagger}(g,\mathbf{k})
Z(R \mathbf{k})
\tilde{d}(g,\mathbf{k})
\Bigr ]_{\mu \mu'}.
\end{align}
Here $Z(R \mathbf{k})$ is the projection operator
defined as\cite{Souza01}
\begin{equation}
Z_{\mu \mu'}(\mathbf{k})=
\sum_{\mathbf{b}} \textrm{w}_{b}
\sum_{I}
\langle u^{\textrm{KS}}_{\mathbf{k}\mu}| u^{(opt)}_{\mathbf{k}+\mathbf{b} I} \rangle
\langle u^{(opt)}_{\mathbf{k}+\mathbf{b} I}| u^{\textrm{KS}}_{\mathbf{k}\mu'} \rangle,
\label{eq:Zmat}
\end{equation}
and $\lambda_{I'I}(\mathbf{k})$ is calculated as
$\lambda_{I'I}(\mathbf{k})=\langle u^{(opt)}_{\mathbf{k} I'}|
\tilde{Z}(\mathbf{k})
|u^{(opt)}_{\mathbf{k} I} \rangle$.
In practice, for each state $I$, we diagonalize
$\tilde{Z}(\mathbf{k})$ in the subspace spanned by
$\psi^{(opt)}_{\mathbf{k} I}$ and
$\delta \psi^{(opt)}_{\mathbf{k} I}$,
and we construct a new $\psi^{(opt)}_{\mathbf{k} I}$ from
the eigenvector with the larger eigenvalue of this $2 \times 2$ matrix. In each iteration,
after all $\psi^{(opt)}_{\mathbf{k}I}$ are updated, we
orthonormalize $U_{\mu I}^{(opt)}(\mathbf{k})$
and impose the condition Eq.~(\ref{eq:Uk}) by using the method
described above.
This point is an important difference between the current
scheme and the usual maximally-localized Wannier function
approach; in the latter
$U^{(opt)}_{\mu I}(\mathbf{k})$ is chosen to be the eigenvectors
of the $N$ largest
eigenvalues of $Z(\mathbf{k})$ (Eq.~(\ref{eq:Zmat})).
The optimal subspace chosen in the conventional
approach
does not necessarily match the subspace spanned by
the desired symmetry-adapted Wannier functions.

After the wavefunctions
$\psi^{(opt)}_{\mathbf{k} I}(\mathbf{r})$ are obtained, we calculate
the transformation matrix $\tilde{U}_{I' I}(\mathbf{k})$
that yields the Bloch wavefunctions(Eq.~(\ref{eq:qBloch})) as a linear combination of
$\psi^{(opt)}_{\mathbf{k} I}(\mathbf{r})$,
\begin{equation}
\psi_{\mathbf{k}I}(\mathbf{r})=
\sum_{I'} \tilde{U}_{I' I}(\mathbf{k})
\psi^{(opt)}_{\mathbf{k}I'}(\mathbf{r})
\end{equation}
which yields
the minimum of $\Omega_{D}+\Omega_{OD}$ in the
chosen subspace, in
the same way as in the isolated band case.
Since $\psi_{\mathbf{k} I}(\mathbf{r})$ and
$\psi^{(opt)}_{\mathbf{k} I}(\mathbf{r})$ both transform
according to Eq.~(\ref{eq:Bloch}), the relation of
the transformation matrices $\tilde{U}(R\mathbf{k})$ and
$\tilde{U}(\mathbf{k})$
is modified from Eqs.~(\ref{eq:URk}) and (\ref{eq:Uk}) as follows:
\begin{equation}
\tilde{U}(R\mathbf{k})D(g,\mathbf{k})=
D(g,\mathbf{k})\tilde{U}(\mathbf{k}),
\end{equation}
\begin{align}
\tilde{U}(\mathbf{k})&=
D(g_{\mathbf{k}},\mathbf{k})\tilde{U}(\mathbf{k})
D^{\dagger}(g_{\mathbf{k}},\mathbf{k})
\\
& 
\qquad\qquad
(\mathbf{k}\in \textrm{IBZ}, g_{\mathbf{k}}\in G_{\mathbf{k}}).
\nonumber
\end{align}
\subsection{Computational details}
In this work we perform calculations using
 the plane-wave DFT code TAPP\cite{Yamauchi96} with
norm-conserving Troullier-Martins type
pseudopotentials\cite{Troullier91}.
We employ
the generalized gradient approximation\cite{Perdew96}
for the exchange-correlation functional.
For the minimization of the spread functional,
the routines in Wannier90 library\cite{wannier90} are used.
All calculations are done using experimental lattice 
constants, that are $a=5.65 \textrm{\AA}$ and $a=3.61 \textrm{\AA}$
for GaAs and Cu, respectively,
and we use
$4 \times 4 \times 4$ and
$8 \times 8 \times 8$
k-point sampling including $\mathbf{k}=\mathbf{0}$ 
(the $\Gamma$ point).
Energy cutoffs of the plane-wave basis
are 25 Ry and 64 Ry for GaAs and Cu,
respectively. Spin-orbit coupling is not included in the calculations.
\section{results}
\subsection{GaAs}
\begin{figure}[tbp]
\includegraphics[clip]{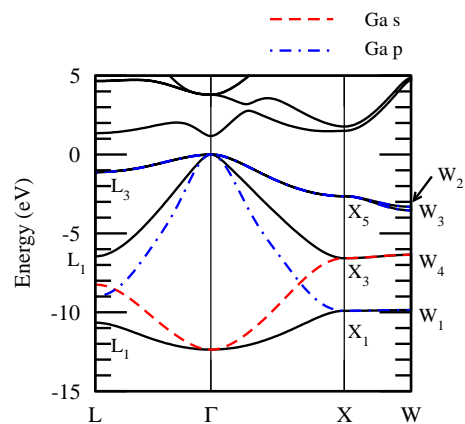}
\caption{(Color online) Interpolated band structure
of GaAs obtained from
the Ga-centered $s$-like Wannier function(dashed lines)
and the Ga-centered
$p$-like Wannier functions(dash-dotted lines).
%calculated separately
%from the four valence bands.
The solid lines show the original band structure.}
\label{fig:gaasband}
\end{figure}

%% \begin{figure}[tbp]
%% \includegraphics[clip]{gaasband.eps}
%% \caption{Band structure of GaAs.}
%% \label{fig:gaasband}
%% \end{figure}
First we consider constructing
four Wannier functions from the four valence
bands in GaAs, whose band structure is shown in the solid lines in Fig.~\ref{fig:gaasband}.
As shown in Ref.~\onlinecite{Marzari97},
in this system
the maximal localization procedure yields four localized functions
centered on four covalent bonds. From group-theoretical view,
those correspond to
the irreducible representation $a_{1g}$ of site-symmetry group of
the Wyckoff position $e$, and one can obtain the same results with
our symmetry-constrained minimization procedure.
In this system
any point along the bond yields the same set of the
matrices $D(g,\mathbf{k})$ (Eq.~(\ref{eq:DRk})), therefore
starting from the initial Wannier functions
centered at an arbitrary point along the bond 
and its symmetry-equivalent three points,
after the minimization
their centers are moved to the points
which yield the minimum of the spread
functional, that are around
$0.155 \times \sqrt{3} a$ away from
Ga atom.

Another set of symmetry-adopted Wannier functions that are
compatible with the symmetry of these four valence states are
$s$-like and $p$-like functions centered at the anion(As) atom,
which correspond to the irreducible representations
$a_{1}$ and $t_{2}$, respectively,
of site-symmetry group of the
Wyckoff position $c (\frac{1}{4}\frac{1}{4}\frac{1}{4})$.
In the usual minimization without symmetry constraint,
these atom-centered Wannier functions are a stationary point
of the spread functional but not the global minimum of it, and
therefore they are not a stable solution
as discussed by Marzari and Vanderbilt\cite{Marzari97}.
In our approach, these atom-centered Wannier functions are
easily obtained by providing corresponding
matrix $D(g,\mathbf{k})$.
In Table \ref{tbl:gaasomega} we compare the spreads
of the bond-centered and atom-centered Wannier functions.
Following Ref.~\onlinecite{Marzari97},
in the table we also show the results obtained
by combining two independent calculations
for $s$- and $p$-like Wannier functions; namely,
in this calculation
we first perform two calculations to obtain
the $s$-like and $p$-like Wannier functions separately, from
the lowest band and higher three bands, respectively.
By using these $1\times 1$ and $3\times 3$ transformation matrices
we construct the $4 \times 4$ transformation matrix $U_{\mu I}(\mathbf{k})$
in the block-diagonal form
without further optimization.
As anticipated,
compared to this separate result,
the atom-centered Wannier functions constructed
with four valence bands are more localized. This is mainly
due to the reduction in the off-diagonal contribution of the
spread functional (Eq.~(\ref{eq:Omega_OD})).

In this system
it is also possible to construct $s$-like
and $p$- functions centered
at the cation(Ga) atom from the four valence bands.
These $s$- and $p$-like functions
correspond to the irreducible representations
$a_{1}$ and $t_{2}$, respectively,
of site-symmetry group of Wyckoff position $a(000)$.
The spreads of these cation-centered functions
are also shown in Table \ref{tbl:gaasomega}, and
as anticipated, these Wannier functions are more delocalized
compared to bond-centered or anion-centered ones.
Unlike the As-centered case, it is not possible to construct
these $s$- and $p$-like Ga-centered functions
separately from the lowest
band and other three bands;
Equation (\ref{eq:Uk}) cannot be fulfilled separately for
$1\times 1$ and $3 \times 3$ unitary transformation
matrices for the lowest band and higher three bands,
respectively, but it can be fulfilled if
we use the four valence bands together.
The reason for this becomes
clear from Fig.~\ref{fig:gaasband}, where we plot
the Wannier-interpolated band-structure\cite{Souza01}
calculated separately from
the $s$-like Wannier function and from
the three $p$-like Ga-centered Wannier functions.
One can see the Ga-centered $s$-like Wannier function
is connected to $X_{3}$ and $W_{3}$ states,
not the lowest $X_{1}$ and $W_{1}$ states which are
constructed from the $p$-like functions.
This shows a close connection
between the symmetry of the Wannier functions and the band structure;
the correspondence between irreducible
representations of a given
site symmetry group and the Bloch functions at high-symmetry
$\mathbf{k}$-points can be 
found in the tables in Ref.~\onlinecite{Evarestov97}.
At the $L$ point,
there are two states belonging to
the same irreducible representation ($L_{1}$) that
contribute to both the $s$-like and $p$-like Wannier functions,
therefore at this point
the two interpolated bands deviate from
the original ones.
We finally note that
the original band structure of the four valence bands
can be reproduced
by using these $s$-like
and $p$-like Ga-centered
Wannier functions together in the interpolation.

\begin{table}%[htb]
\begin{ruledtabular}
\begin{tabular}{r c c c c c c}
    &
$\Omega_{\textrm{I}}$  &
$\Omega_{\textrm{D}}$  &
$\Omega_{\textrm{OD}}$ &
$\Omega$            &
\multicolumn{2}{c}{$\Omega_{n}$}          \\ \hline
bond-centered & & & & & & \\
$4 \times 4 \times 4$ &
6.124 & 0.006 & 0.630 & 6.760 & \multicolumn{2}{c}{1.690} \\
$8 \times 8 \times 8$ &
7.870 & 0.006 & 0.566 & 8.442 & \multicolumn{2}{c}{2.110} \\
 centered on As & & & & & $s$ & $p$\\
$4 \times 4 \times 4$ &
6.124 & 0.012 & 3.502 & 9.639 & 1.450 & 2.730 \\
$8 \times 8 \times 8$ &
7.870 & 0.012 & 3.826 & 11.708 & 1.510 & 3.399 \\
*centered on As & & & & & $s$ & $p$\\
$4 \times 4 \times 4$ &
6.124 & 0.064 & 4.388 & 10.576 & 1.828 & 2.916 \\
$8 \times 8 \times 8$ &
7.870 & 0.069 & 4.943 & 12.882 & 2.032 & 3.617 \\
 centered on Ga & & & & & $s$ & $p$\\
$4 \times 4 \times 4$ &
6.124 & 0.151 & 7.648 & 13.924 & 2.448 & 3.825 \\
$8 \times 8 \times 8$ &
7.870 & 0.112 & 9.028 & 17.011 & 2.615 & 4.798 \\
\end{tabular}
\caption{Spreads of the four
bond-centered and atom-centered Wannier functions 
of GaAs in \AA$^{2}$
calculated
with $4 \times 4 \times 4$ and
$8 \times 8 \times 8$ $\mathbf{k}$-point sampling. The asterisk
(*) shows the results obtained by combining
the solutions of separate one-band
and three-band calculations as done
in Ref.~\onlinecite{Marzari97},
and $\Omega_{n}$ denotes the
spread of one Wannier function.}
\label{tbl:gaasomega}
\end{ruledtabular}
\end{table}

\subsection{Cu}
\begingroup
\squeezetable
\begin{table}%[htb]
\begin{ruledtabular}
\begin{tabular}{r c c c c c c c}
    &
$\Omega_{\textrm{I}}$  &
$\Omega_{\textrm{D}}$  &
$\Omega_{\textrm{OD}}$ &
$\Omega$            &
\multicolumn{3}{c}{$\Omega_{n}$}          \\ \hline
\multicolumn{8}{c}{Energy window [-10 eV :10 eV]}  \\
atom-centered     & & & & & $s$ & $t_{2g}$ & $e_{g}$ \\
$4 \times 4 \times 4$ &
3.901 & 0.000 & 0.248 & 4.149 & 1.959 & 0.439 & 0.437 \\
$8 \times 8 \times 8$ &
5.613 & 0.000 & 0.208 & 5.820 & 3.474 & 0.466 & 0.475 \\
centered at
$(\frac{1}{2}\frac{1}{2}\frac{1}{2})$ &
 & & & & $s$ & $t_{2g}$ & $e_{g}$ \\
$4 \times 4 \times 4$ &
3.555 & 0.000 & 0.407 & 3.962 & 1.706 & 0.447 & 0.457 \\
$8 \times 8 \times 8$ &
4.815 & 0.000 & 0.561 & 5.376 & 2.855 & 0.488 & 0.528 \\
centered at
$(\frac{1}{4}\frac{1}{4}\frac{1}{4})$ &
 & & & & $s$ & $t_{2g}$ & $e_{g}$ \\
$4 \times 4 \times 4$ &
3.279 & 0.167 & 0.500 & 3.946 & 1.599 & 0.489 & 0.441 \\
$8 \times 8 \times 8$ &
3.968 & 0.107 & 0.511 & 4.587 & 2.042 & 0.534 & 0.471 \\
\multicolumn{8}{c}{Energy window [-10 eV :20 eV]}  \\
atom-centered     & & & & & $s$ & $t_{2g}$ & $e_{g}$ \\
$4 \times 4 \times 4$ &
3.342 & 0.000 & 0.041 & 3.382 & 1.463 & 0.386 & 0.381 \\
$8 \times 8 \times 8$ &
3.685 & 0.000 & 0.016 & 3.701 & 1.705 & 0.400 & 0.397 \\
centered at
$(\frac{1}{2}\frac{1}{2}\frac{1}{2})$ &
 & & & & $s$ & $t_{2g}$ & $e_{g}$ \\
$4 \times 4 \times 4$ &
3.059 & 0.000 & 0.164 & 3.222 & 1.247 & 0.393 & 0.398 \\
$8 \times 8 \times 8$ &
3.347 & 0.000 & 0.176 & 3.523 & 1.439 & 0.412 & 0.424 \\
centered at
$(\frac{1}{4}\frac{1}{4}\frac{1}{4})$ &
 & & & & $s$ & $t_{2g}$ & $e_{g}$ \\
$4 \times 4 \times 4$ &
2.947 & 0.011 & 0.250 & 3.208 & 1.164 & 0.422 & 0.390 \\
$8 \times 8 \times 8$ &
3.168 & 0.008 & 0.252 & 3.428 & 1.284 & 0.443 & 0.407 \\
\end{tabular}
\caption{Spreads of the Wannier functions of Cu in \AA$^{2}$ 
calculated
with $4 \times 4 \times 4$ and
$8 \times 8 \times 8$ $\mathbf{k}$-point sampling.
The labels refer to the center of the $s$-like function,
and the
$t_{2g}$ and $e_{g}$ functions are always on the atom.
$\Omega_{n}$ denotes the
spread of one Wannier function.}
\label{tbl:cusixomega}
\end{ruledtabular}
\end{table}
\endgroup

%% \begingroup
%% \squeezetable
%% \begin{table}%[htb]
%% \begin{ruledtabular}
%% \begin{tabular}{r c c c c c c c c}
%%     &
%% $\Omega_{\textrm{I}}$  &
%% $\Omega_{\textrm{D}}$  &
%% $\Omega_{\textrm{OD}}$ &
%% $\Omega$            &
%% \multicolumn{4}{c}{$\Omega_{n}$}          \\ \hline
%% atom-centered     & & & & & $s$ & $p$ & $t_{2g}$ & $e_{g}$ \\
%% $4 \times 4 \times 4$ &
%% 4.733 & 0.000 & 3.422 & 8.155 & 1.287 & 1.665 & 0.372 & 0.379 \\
%% $8 \times 8 \times 8$ &
%% 5.628 & 0.000 & 3.888 & 9.516 & 1.434 & 2.048 & 0.385 & 0.391 \\
%% octahedral &
%%  & & & & \multicolumn{2}{c}{$o$} & \multicolumn{2}{c}{$t_{2g}$}  \\
%% $4 \times 4 \times 4$ &
%% 4.733 & 0.010 & 1.825 & 6.568 & \multicolumn{2}{c}{0.909} & \multicolumn{2}{c}{0.372}  \\
%% $8 \times 8 \times 8$ &
%% 5.628 & 0.011 & 1.807 & 7.445 & \multicolumn{2}{c}{1.048} & \multicolumn{2}{c}{0.385}  \\
%% \label{tbl:cunineomega}
%% \end{tabular}
%% \caption{Spreads of Cu in \AA$^{2}$ 
%% calculated
%% with $4 \times 4 \times 4$ and
%% $8 \times 8 \times 8$ $\mathbf{k}$-point sampling.
%% $\Omega_{n}$ denotes the
%% spread of one Wannier function.}
%% \end{ruledtabular}
%% \end{table}
%% \endgroup
Next we consider 
constructing six Wannier functions from one $s$-like
and five $d$-like states for bulk copper in fcc structure.
Souza \textit{et al.}\cite{Souza01} showed that in this system
one obtains five $d$-like Wannier functions centered
on Cu atom
that
are split into $t_{2g}$ and $e_{g}$ states
and
one $s$-like Wannier function whose
center is not on Cu atom
but at the tetrahedral interstitial site
$(\frac{1}{4} \frac{1}{4} \frac{1}{4})$.
In the six-band case,
this tetrahedrally-centered Wannier function
is not regarded as a symmetry-adapted Wannier function, as
due to the inversion symmetry this site
(Wyckoff position $c$) is equivalent to
$(\frac{-1}{4}\frac{-1}{4}\frac{-1}{4})$ and thus 
one needs one additional $s$-like Wannier function centered
at the latter site
to make them
the basis functions of the full symmetry group,
resulting a seven-band model as discussed by
Souza \textit{et al}\cite{Souza01}.

The possible $s$-like symmetry-adapted Wannier function
in this six-band case
is (i) $a_{1g}$ irreducible representation centered on Cu
atom(Wyckoff position $a$) and (ii)
$a_{1g}$ irreducible representation
centered at $(\frac{1}{2}\frac{1}{2}\frac{1}{2})$
(Wyckoff position $b$).
In Table \ref{tbl:cusixomega}, we compare
the spreads of these sets of the six
Wannier functions
obtained by using the energy window of [-10eV:+10eV] and
[-10eV:+20eV]. In these calculations,
the centers of the $t_{2g}$ and $e_{g}$ Wannier
functions are on the Cu atom.
As expected,
the two symmetry-adapted solutions
yield larger spreads than the tetrahedrally-centered
Wannier functions obtained via the unconstrained minimization,
while the spreads of $d$-like Wannier functions do not
vary very much in these three cases.
The $s$-like Wannier function centered at
$(\frac{1}{2}\frac{1}{2}\frac{1}{2})$ is found to
be more localized than
the atom-centered Wannier function, which may be traced back to
the fact that the $s$-like band
in Cu is very extended and it has a larger weight
in the interstitial region.

\begin{figure}[tbp]
\includegraphics[clip]{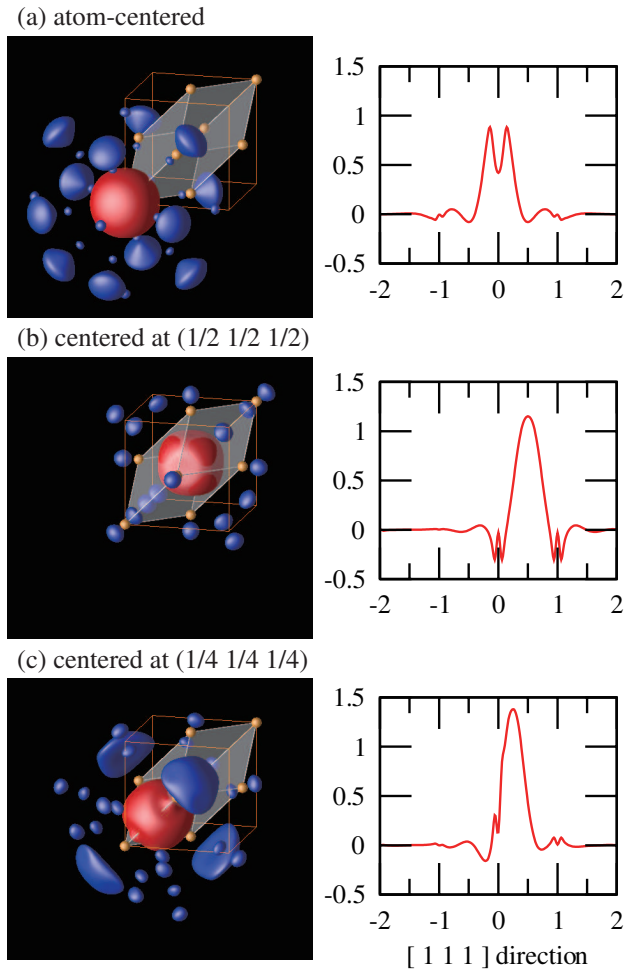}
\caption{(Color online) The $s$-like
Wannier functions of Cu for three cases
calculated with the energy window of [-10eV:+10eV].
Left: Isosurfaces at $+0.75/\sqrt{V}$ and
$-0.125/\sqrt{V}$, where $V$ is the volume of the unit cell. The Cu atoms
in the unit cell are also shown by spheres.
Right:
Plots along $[111]$ direction.
The unit of the horizontal axes is
$\sqrt{3}a$.}
\label{fig:cuwandx}
\end{figure}

The gauge-invariant part of the spread functional
($\Omega_{\textrm{I}}$, Eq.~(\ref{eq:Omega_I})) is also different in the three cases,
which indicates that
the optimal subspace(Eq.~(\ref{eq:psiopt}))
chosen in the first step of the minimization procedure 
is different in these three cases as
a result of the symmetry constraint.
As reported previously
by Souza \textit{et al}.\cite{Souza01} and
Thygesen \textit{et al.}\cite{Thygesen05},
in the case of the energy window [-10eV:+10eV],
we find
that without symmetry constraint the atom-centered
$s$-like Wannier function is not a stable solution.
In the case of the larger
 energy window of [-10eV:+20eV], we get the atom-centered
$s$-like Wannier function without symmetry constraint
by using atom-centered gaussian functions
as initial trial functions,
however we find that without symmetry constraint
this solution is unstable against a small
perturbation of the initial states,
and this clearly shows
the importance of the symmetry constraint
when constructing Wannier functions from extended bands.
In Fig.~\ref{fig:cuwandx} we plot these three $s$-like
Wannier functions. 
Since the two symmetry-adapted functions
(Fig.~\ref{fig:cuwandx} (a) and (b)) belong to 
the $a_{1g}$ irreducible representation,
they are invariant with respect to transformations of
their site symmetry group. 
It can be seen that 
the most localized tetrahedrally-centered
solution (Fig.~\ref{fig:cuwandx} (c)),
which is obtained without
symmetry constraint, is also symmetric with respect to
the rotation of $\frac{2\pi}{3}$ around the [111] axis,
which indicates this function also reflects
some site-symmetry properties of the tetrahedral site.

%\begin{figure}[tbp]
%\includegraphics[clip]{cuwan111.eps}
%\caption{(Color online) Wannier functions of the
%$s$-like state of Cu along $[1 1 1]$ direction.}
%\label{fig:cuwan111}
%\end{figure}

\begin{figure}[tbp]
\includegraphics[clip]{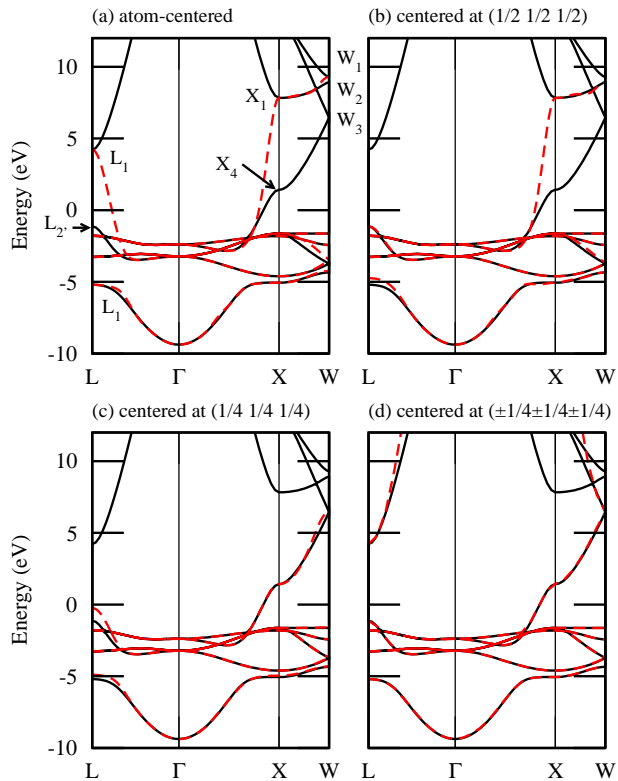}
\caption{(Color online) LDA band structure (solid lines) and
Wannier interpolated band structure (dashed lines) of Cu calculated from
different set of Wannier functions:
(a) atom-centered $sd^{5}$ functions.
(b) five atom-centered
$d$ functions and one $s$-like function centered
at $(\frac{1}{2} \frac{1}{2} \frac{1}{2})$.
(c) five atom-centered
$d$ functions and one $s$-like function centered
at $(\frac{1}{4} \frac{1}{4} \frac{1}{4})$ (broken-symmetry solution).
(d)  five atom-centered
$d$ functions and two $s$-like functions centered
at $(\pm\frac{1}{4} \pm\frac{1}{4} \pm\frac{1}{4})$. In Fig.~\ref{fig:cuband}(d)
the inner (frozen) window\cite{Souza01} of [-10eV:7.2eV] is used.
}
\label{fig:cuband}
\end{figure}

%Finally we discuss the relation between
%the Wannier function and band structure.
In Fig.~\ref{fig:cuband} we plot Wannier-interpolated
band structures calculated with
different sets of the Wannier functions.
For completeness,
in Fig.~\ref{fig:cuband}(d)
we also show the interpolated band
structure calculated with seven Wannier functions, namely,
five $d$-like Wannier functions and two equivalent
$s$-like Wannier functions centered at
($\frac{\pm 1}{4}\frac{\pm 1}{4}\frac{\pm 1}{4}$).
As in the case of GaAs and
as also discussed by Souza \textit{et al.}\cite{Souza01},
for high-symmetry points in the Brillouin zone
one can predict which
Bloch states can be formed from a given set of the symmetry-adapted
Wannier functions; as seen in Fig.~\ref{fig:cuband}
the atom-centered $s$-like
Wannier function (Fig.~\ref{fig:cuband}(a))
is connected
to $L_{1}$, $X_{1}$, and $W_{1}$ states, while
from the Wannier function centered
at $(\frac{1}{2}\frac{1}{2}\frac{1}{2})$,
$L_{2'}$, $X_{1}$, and $W_{2}$ states are formed.
The low-lying $X_{4}$ and $W_{3}$ states are formed
with the tetrahedrally-centered Wannier functions
(Figs.~\ref{fig:cuband}(c) and (d)),
as discussed by Souza \textit{et al}\cite{Souza01}.

As can be seen in Fig.~\ref{fig:cuband}(a),
the $L_{1}$, $X_{1}$, and $W_{1}$ states which are
formed by the atom-centered $s$-like Wannier function
are located in a relatively high energy region,
and this is why the atom-centered $s$-like Wannier function
is unstable in the conventional maximal localization
approach when a small energy window is chosen.
Indeed, in our calculation, with a smaller choice of
the energy window, we cannot satisfy the relation
Eq.~(\ref{eq:Uk}) for
the atom-centered $s$-like Wannier function.
This shows the importance of 
selecting the energy window properly,
as the symmetry of the Wannier functions are
determined by the symmetry properties
of the Bloch functions inside the energy
window through Eqs.~(\ref{eq:URk}) and (\ref{eq:dks}).

\section{conclusions}
In this paper we have presented a systematic
procedure to generate
symmetry-adapted Wannier functions based on the theory of
site-symmetry and induction group combined with the maximally-localized Wannier function approach. This scheme can easily
be implemented in the existing maximally-localized Wannier
function calculation code,
and it allows
one to calculate localized functions of a specified symmetry
which do not necessarily yield the global minimum of the
spread functional.
It also provides the relation between
the unitary transformation
matrices for symmetry-equivalent $\mathbf{k}$-points,
which simplifies the minimization process
 and also improves accuracy of the calculation.

The results for GaAs and Cu show that the
calculated Wannier functions are indeed localized
and have the specified symmetry properties, and they reflect
the symmetry of the Bloch functions inside the energy
window used in the calculation.

These symmetry-adapted Wannier functions are suitable
for symmetry analysis of the band-structure of the
system and for
accurate basis functions of the tight-binding model.
Generalizations of the present method, such as
including spin-orbit coupling,
would be interesting subjects to be investigated.

\begin{acknowledgments}
We thank K. Nakamura for providing us with
his pseudopotential data, and F. Aryasetiawan and
F. Nilsson for helpful discussions.
This work was supported by Swedish Research Council
and the Scandinavia-Japan Sasakawa Foundation.
\end{acknowledgments}

\end{document}